\documentclass[
 reprint,
 amsmath,amssymb,
 aps,prb,superscriptaddress
]{revtex4-1}

\usepackage{graphicx}
\usepackage{amsmath}
\usepackage{bm}
\usepackage{hyperref}
\usepackage{tabularx}
\usepackage[version=4]{mhchem} 
\hypersetup{
    colorlinks=true,
    linkcolor=blue,
    filecolor=magenta,      
    urlcolor=blue,
    citecolor=blue
}

\begin{document}

\title{Efficient lattice dynamics calculations for correlated materials with DFT+DMFT}
\author{Can P. Ko\c{c}er}
\email{cpk27@cam.ac.uk}
\affiliation{Theory of Condensed Matter, Cavendish Laboratory, University of Cambridge, J. J. Thomson Avenue, Cambridge CB3 0HE, UK}
\author{Kristjan Haule}
\affiliation{Department of Physics and Astronomy, Rutgers University, Piscataway, New Jersey 08854, USA}
\author{G. Lucian Pascut}
\affiliation{MANSiD Research Center and Faculty of Forestry, Applied Ecology Laboratory, Stefan Cel Mare University (USV), Suceava 720229, Romania}
\author{Bartomeu Monserrat}
\email{bm418@cam.ac.uk}
\affiliation{Theory of Condensed Matter, Cavendish Laboratory, University of Cambridge, J. J. Thomson Avenue, Cambridge CB3 0HE, UK}
\affiliation{Department of Materials Science and Metallurgy, University of Cambridge, 27 Charles Babbage Road, Cambridge CB3 0FS, United Kingdom}
\date{\today}

\begin{abstract}
Phonons are fundamentally important for many materials properties, including thermal and electronic transport, superconductivity, and structural stability. Here, we describe a method to compute phonons in correlated materials using state-of-the-art DFT+DMFT calculations. Our approach combines a robust DFT+DMFT implementation to calculate forces with the direct method for lattice dynamics using nondiagonal supercells. The use of nondiagonal instead of diagonal supercells drastically reduces the computational expense associated with the DFT+DMFT calculations. We benchmark the method for typical correlated materials (Fe, NiO, MnO, SrVO\textsubscript{3}), testing for $\mathbf{q}$-point grid convergence and different computational parameters of the DFT+DMFT calculations. The efficiency of the nondiagonal supercell method allows us to access $\mathbf{q}$-point grids of up to $6\times6\times6$. In addition, we discover that for the small displacements that atoms are subject to in the lattice dynamics calculation, fixing the self-energy to that of the equilibrium configuration is in many cases an excellent approximation that further reduces the cost of the DFT+DMFT calculations. This fixed self-energy approximation is expected to hold for materials that are not close to a phase transition. Overall, our work provides an efficient and general method for the calculation of phonons using DFT+DMFT, opening many possibilities for the study of lattice dynamics and associated phenomena in correlated materials.
\end{abstract}

\maketitle

\section{Introduction}
First-principles calculations of phonons in real materials play an important role in explaining experimental observations and in predicting novel materials phenomena. Apart from standalone calculations, lattice dynamics often form the input for a wide variety of follow-up calculations of materials properties, including thermodynamics, superconductivity, thermal and electronic transport, and finite temperature optical response.

The overwhelming majority of lattice dynamics calculations of materials employ density functional theory (DFT). However, commonly employed exchange-correlation functionals such as LDA and PBE have severe shortcomings when applied to materials with strongly correlated \textit{d} or \textit{f} electrons. While these problems can be partially remedied by DFT+U methods or hybrid functionals, dynamical mean field theory (DMFT) in combination with DFT generally leads to a better description of the electronic structure of correlated materials~\cite{kotliar2006}. It is therefore desirable to extend the range of applicability of DFT+DMFT calculations to study structural and vibrational properties of correlated materials. To this end, DFT+DMFT implementations  for total energies and forces have been developed recently~\cite{haule2015,haule2016,haule2017,leonov2014a}.

Phonon calculations are usually performed by one of three methods: linear response, frozen phonons, or the direct method. In the context of DMFT, early work by Savrasov and Kotliar~\cite{savrasov2003} described a linear response method to calculate phonon spectra of MnO and NiO. The authors used the simple Hubbard-I solver and neglected the change of the self-energy with displacement, a term that involves the derivative of the self-energy $\Sigma$ with respect to the Green's function $G$, $\delta\Sigma/\delta G$. This term is very difficult to compute by the current generation of impurity solvers. The frozen phonon method was used in the work of Leonov \textit{et al.}~\cite{leonov2012} to calculate lattice dynamics of paramagnetic iron and more recently by Appelt \textit{et al.} to compute the phonons of palladium~\cite{appelt2020}. Frozen phonon calculations rely on \textit{a priori} knowledge of the phonon eigenvectors so that phonon frequencies are easily calculated from total energy differences without the need to evaluate forces. As such, the method only applies to simple, highly symmetric structures. The direct method is both simple and general, requiring only the forces on atoms, and given the recent advances in force implementations of DFT+DMFT, should be the method of choice. A recent example of this is the study of phonons in iron by Han \textit{et al.}~\cite{han2018}. However, as it relies on the construction of supercells to access phonons at points other than $\Gamma$, the computational cost can quickly become unmanageable for an already expensive electronic structure method such as DMFT.

In this paper, we describe and benchmark a method to compute vibrational properties of correlated materials from DFT+DMFT. The method combines two ingredients: (1) Forces from DFT+DMFT, efficiently obtained from a robust implementation based on the free-energy Luttinger-Ward functional~\cite{haule2016}, and (2) the direct method for phonon calculations, using non-diagonal rather than diagonal supercells for significant savings in computational expense~\cite{lloyd-williams2015}. In addition, we discover that using a fixed self-energy obtained from the equilibrium configuration for the configurations with atomic displacements is an excellent approximation. Since the solution of the impurity problems is the most expensive step of the calculations, this approximation results in additional large savings of computational time.

The paper is organised as follows: In Sec.~\ref{sec:Methods}, we describe background theory and implementation for the DFT+DMFT method and lattice dynamics calculations with nondiagonal supercells. In Sec.~\ref{sec:Results}, results of the calculations for Fe, NiO, MnO and \ce{SrVO3} are described and discussed in turn, including tests for $\mathbf{q}$-point grid convergence, use of fixed self-energies, and other computational parameters. We draw conclusions and outline future work in Sec.~\ref{sec:Conclusion}.

\section{Methods}\label{sec:Methods}

\subsection{DFT+DMFT}
The DFT+DMFT calculations are based on the method and implementation of Haule \textit{et al.}~\cite{haule2010,haule,haule2015,haule2018}, often referred to as DFT + embedded DMFT (DFT+eDMFT). In this method, the DFT+DMFT free energy is expressed in the form of a Luttinger-Ward functional, which is stationary. This stationarity is important as it allows reliable evaluation of free energies and forces~\cite{haule2016}. To connect the correlated subspaces to the rest of the solid, projection operators $\hat{P}^{\mathbf{R}}$ are defined such that $G^{\mathbf{R}}_\mathrm{loc}=\hat{P}^{\mathbf{R}}G$, where $G$ and $G^{\mathbf{R}}_{\mathrm{loc}}$ are the Green's function of the solid and the local Green's function of the correlated atom at site $\mathbf{R}$, respectively. On-site correlations of $d$ or $f$ orbitals are treated exactly while more itinerant degrees of freedom are treated on the DFT level. The projectors are fixed and consist of a set of quasi-atomic orbitals $\phi^{\mathbf{R}}_{lm}(\mathbf{r})$ that are solution to the Schr\"odinger equation inside the muffin-tin sphere. The projection and embedding with fixed projectors is required to preserve the stationary nature of the functional. The DFT+DMFT calculation proceeds as follows: (1)~the Green's function of the lattice $G$ is projected to the local orbital basis ($d$ or $f$ orbitals) to calculate the local Green's functions $G^{\mathbf{R}}_{\mathrm{loc}}$ for each independent correlated atom at sites $\{\mathbf{R}\}$, (2)~the impurity problem for each independent correlated atom is solved using the continuous time quantum monte carlo~\cite{werner2006,haule2007} (CTQMC) solver to obtain the self-energy in the local orbital basis $\Sigma_{\alpha\beta}(\omega)$, (3)~the self-energy is embedded into real space according to
\begin{equation}\label{eqn:selfenergy}
    \Sigma^{\mathbf{R}}(\mathbf{r},\mathbf{r}',\omega) = \sum_{\alpha,\beta} \langle \mathbf{r} \vert \phi_\alpha \rangle \Sigma_{\alpha\beta}(\omega) \langle \phi_\beta \vert \mathbf{r}' \rangle
\end{equation}
and is nonzero only within the muffin-tin spheres of the correlated atoms. The self energy then enters the Dyson equation of the solid to obtain the lattice Green's function. Self-consistency is achieved when the local Green's functions obtained from lattice and impurity match.

The forces on atoms are defined as the derivatives of the Luttinger-Ward free energy functional with respect to the atomic positions, which includes the effects of electronic and magnetic entropy~\cite{haule2016}. Importantly, they are easily and reliably evaluated in this implementation, being even more numerically precise to compute than the free energy. Accurate forces are essential for calculating phonons from finite differences. We note that other implementations of forces within DFT+DMFT exist; for example the work of Leonov \textit{et al.} (Ref.~\cite{leonov2014a}). In contrast to the force implementation used in our work, the method in Ref.~\cite{leonov2014a} uses a Wannier function basis and does not define the force as the derivative of a stationary free energy functional. Despite these difference, the implementation of Ref.~\cite{leonov2014a} should in principle also be suited for lattice dynamics calculations with the direct method.

All DFT+DMFT calculations were performed using the code available at~\cite{haule}. The DFT part of the calculation is based on the \texttt{WIEN2K} code~\cite{blaha2018}, using an all-electron LAPW basis set. The LDA is used throughout as the exchange-correlation functional for the DFT part. A window of 20~eV around the Fermi level is used for the hybridisation. The DMFT calculations were performed using the exact double counting~\cite{haule2015a}. Experimental lattice parameters were obtained from Ref.~\cite{basinski1955} for Fe, Ref.~\cite{sasaki1979} for MnO and NiO, and Ref.~\cite{lan2003} for \ce{SrVO3}. The interaction parameters for Fe were obtained from a previous study that performed constrained DMFT calculations ($U=5.5$~eV, $J=0.84$~eV)~\cite{han2018}. For both MnO and NiO, $U=9.0$~eV was chosen for the correlated $d$-orbitals with $J_{\mathrm{Mn}}=1.14$~eV and $J_{\mathrm{Ni}}=1.3$~eV. For \ce{SrVO3}, $U=6.0$~eV and $J=1.0$~eV were used. Fine tuning of the parameters is avoided in this study. Calculations for the primitive cells of Fe, MnO, and NiO used $\mathbf{k}$-point grids of size $12\times12\times12$, and for \ce{SrVO3} a $10\times10\times10$ grid was used. Equally dense grids were used for all supercell calculations.

\subsection{Lattice Dynamics}
The objective of lattice dynamics calculations in the harmonic approximation is to determine the dynamical matrix at a given $\mathbf{q}$-point in the irreducible Brilouin zone. For a crystal with a primitive cell of $i=1,...,N$ atoms at position $\{\bm{\tau}_i\}$, the dynamical matrix at point $\mathbf{q}$ is defined as
\begin{equation}\label{eqn:dynmat}
D_{i\alpha,j\beta}(\mathbf{q}) = \frac{1}{\sqrt{m_i m_j}}\sum_{\mathbf{R}} \Phi_{i\alpha,j\beta}(\mathbf{R}) e^{i\mathbf{q}\cdot(\mathbf{R}+\bm{\tau}_j-\bm{\tau}_i)}
\end{equation}
where $\alpha,\beta$ label cartesian coordinates and $i,j$ label the atoms within a primitive cell. The masses of the atoms are given by $m_i$ and $m_j$, and
\begin{equation}\label{eqn:ifcmatrix}
\Phi_{i\alpha,j\beta}(\mathbf{R}-\mathbf{R}') = \frac{\partial^2E}{\partial u_{i\alpha\mathbf{R}'} \partial u_{j\beta\mathbf{R}}} = -\frac{\partial F_{j\beta\mathbf{R}}}{\partial u_{i\alpha\mathbf{R}'}}
\end{equation}
is the matrix of interatomic force constants, which is a function of $\mathbf{R}-\mathbf{R}'$ only, due to the translational invariance of the solid. The Born-Oppenheimer potential energy surface $E$ is in this case given by the free energy as obtained from the Luttinger-Ward functional of DFT+DMFT, while $u_{i\alpha\mathbf{R}'}$ corresponds to the displacement of atom $i$ in the primitive cell at positions $\mathbf{R}'$ along cartesian direction $\alpha$. In practice $\Phi_{i\alpha,j\beta}$ is computed row by row through the derivatives of the forces $F_{j\beta\mathbf{R}}$. The phonon frequencies and eigenvectors are found by diagonalising the dynamical matrix. We have used atomic displacements of 0.02~bohr throughout, but tests with 0.01--0.04~bohr show almost identical results for Fe and NiO.

In the direct method of lattice dynamics calculations, the force constant matrix is determined by constructing a supercell. Conventionally, to determine phonon frequencies and eigenvectors on a $N_1\times N_2\times N_3$ $\mathbf{q}$-point grid, a supercell of dimensions $N_1\times N_2\times N_3$ would be constructed. In this work, we use the non-diagonal supercell method of Monserrat \textit{et al.}~\cite{lloyd-williams2015}, which allows a more efficient determination of the force constant and dynamical matrices than the use of a diagonal supercell. The method relies on the fact that a perturbation of the atomic positions that has a wavevector $\mathbf{q}$ is commensurate with a supercell for which $\mathbf{q}$ is a reciprocal lattice vector. It can then be shown that for a $N_1\times N_2\times N_3$ $\mathbf{q}$-point grid, a set of supercells, each of which contains at most a number of primitive cells equal to the least common multiple of $N_1,N_2,N_3$, are sufficient to determine the dynamical matrix at every $\mathbf{q}$-point in the grid. In particular, the method allows the sampling of vibrational Brillouin zones with a uniform grid of size $N\times N\times N$ using supercells that contain at most $N$ primitive cells. In contrast, using only diagonal supercells, the largest of these contains $N^3$ primitive cells. Non-diagonal supercells are solutions to the ``minimum supercell problem" for computing phonons as recently described by Fu \textit{et al.}~\cite{fu2019}, and are the most efficient method (in terms of system size) to compute phonons at a given $\mathbf{q}$-point. The ideas can be generalised to interactions between phonons, as described in Ref.~\cite{fu2019}.

The true force constant matrix satisfies certain sum rules~\cite{ackland1997}. In particular, Newton's third law requires that the sum of the forces on the atoms is zero for every calculation ($\sum_i \mathbf{F}_i = \mathbf{0}$). In terms of the force constant matrix, this means that every row and column must sum to zero:
\begin{equation}
\sum_{j} \Phi_{i\alpha,j\beta} = 0.
\end{equation}
Stated differently, $\Phi_{i\alpha,i\beta}$ must be given by
\begin{equation}\label{eqn:ASR}
\Phi_{i\alpha,i\beta} = -\sum_{j\neq i} \Phi_{i\alpha,j\beta}
\end{equation}
The difference between $\Phi_{i\alpha,i\beta}$ as obtained from the calculation, and calculated by Eq.~\ref{eqn:ASR} can be used as a measure to judge the accuracy or numerical precision of the force evaluations in the \textit{ab initio} calculation~\cite{kunc1982}. This is particularly relevant for DFT+DMFT calculations as the forces are affected by statistical noise. We have also observed that the sum rule and symmetry violations tend to be larger with DFT+DMFT derived forces than for pure DFT calculations. The sum rule and the point group symmetry are therefore applied to the force constant matrices.

In polar insulators, the longitudinal optical (LO) and transverse optical (TO) phonon modes are split close to the $\Gamma$ point due to the interaction between LO phonons and macroscopic electric fields. This LO-TO splitting needs to be taken into account to accurately model the phonon spectra of MnO and NiO. In DFT calculations, LO-TO splitting is included by separately calculating the Born effective charge tensors $Z_i^*$ and the macroscopic dielectric tensor $\epsilon_\infty$, and adding a non-analytic correction to the dynamical matrix~\cite{wang2016a}. Unfortunately, for Green's function based methods like DMFT, the calculations of polarisation (and hence Born effective charges) is still an unsolved theoretical problem. In the limit $\mathbf{q}\to0$, the frequencies of the LO and TO phonon modes $\omega_{\mathrm{LO}}$ and $\omega_{\mathrm{TO}}$ are related to $Z^*$ and $\epsilon_\infty$ according to
\begin{equation}\label{eqn:LOTO}
\omega_{\mathrm{LO}}^2-\omega_{\mathrm{TO}}^2 = \frac{e^2}{\epsilon_0\mu\Omega}\frac{\lvert Z^* \rvert^2}{\epsilon_\infty}
\end{equation}
where $e$ is the elementary charge, $\Omega$ is the volume of the primitive unit cell, $\epsilon_0$ is the vacuum permittivity, and $\mu$ is the reduced mass of the two atoms in the unit cell. The phonon frequencies $\omega_{\mathrm{LO}}$ and $\omega_{\mathrm{TO}}$ can be obtained from a nondiagonal supercell representing a $\mathbf{q}$-point close to $\Gamma$, and the resulting value of $\lvert Z^* \rvert^2/\epsilon_\infty$ is used for the non-analytic correction to the dynamical matrix.

\section{Results and Discussion}\label{sec:Results}

\subsection{Fe}
At ambient pressure, iron crystallises in three different polymorphs: the bcc-$\alpha$ phase (stable below 1185~K), the fcc-$\gamma$ phase (stable between 1185--1670~K), and the bcc-$\delta$ phase (stable up to the melting point of 1811~K). The bcc-$\alpha$ phase is ferromagnetic below the Curie temperature of 1043~K. DFT+DMFT calculations are well suited for the \textit{ab initio} simulation of the interplay between metallicity and local moments in iron~\cite{lichtenstein2001}. Importantly, DFT+DMFT is able to capture both the paramagnetic regime and the temperature-dependent change in the local moment. Within DFT, describing these temperature effects requires much additional work starting from the actual first-principles calculations ~\cite{kormann2012,kormann2014,ikeda2014}.

The temperature dependence of the phonon spectra of elemental iron has been studied previously both experimentally~\cite{neuhaus1997,neuhaus2014} and computationally~\cite{han2018,leonov2012}. In the ferromagnetic bcc $\alpha$-phase, a pronounced softening of the phonon modes is observed as the temperature increases. The phonon softening can be captured with a number of different simulation methods~\cite{kormann2012,kormann2014,ikeda2014}. A recent DFT+DMFT study by Han \textit{et al.} clearly attributed the phonon softening to the melting of the ferromagnetic order~\cite{han2018}.

\begin{figure}[t]
    \centering
    \includegraphics[scale=0.63]{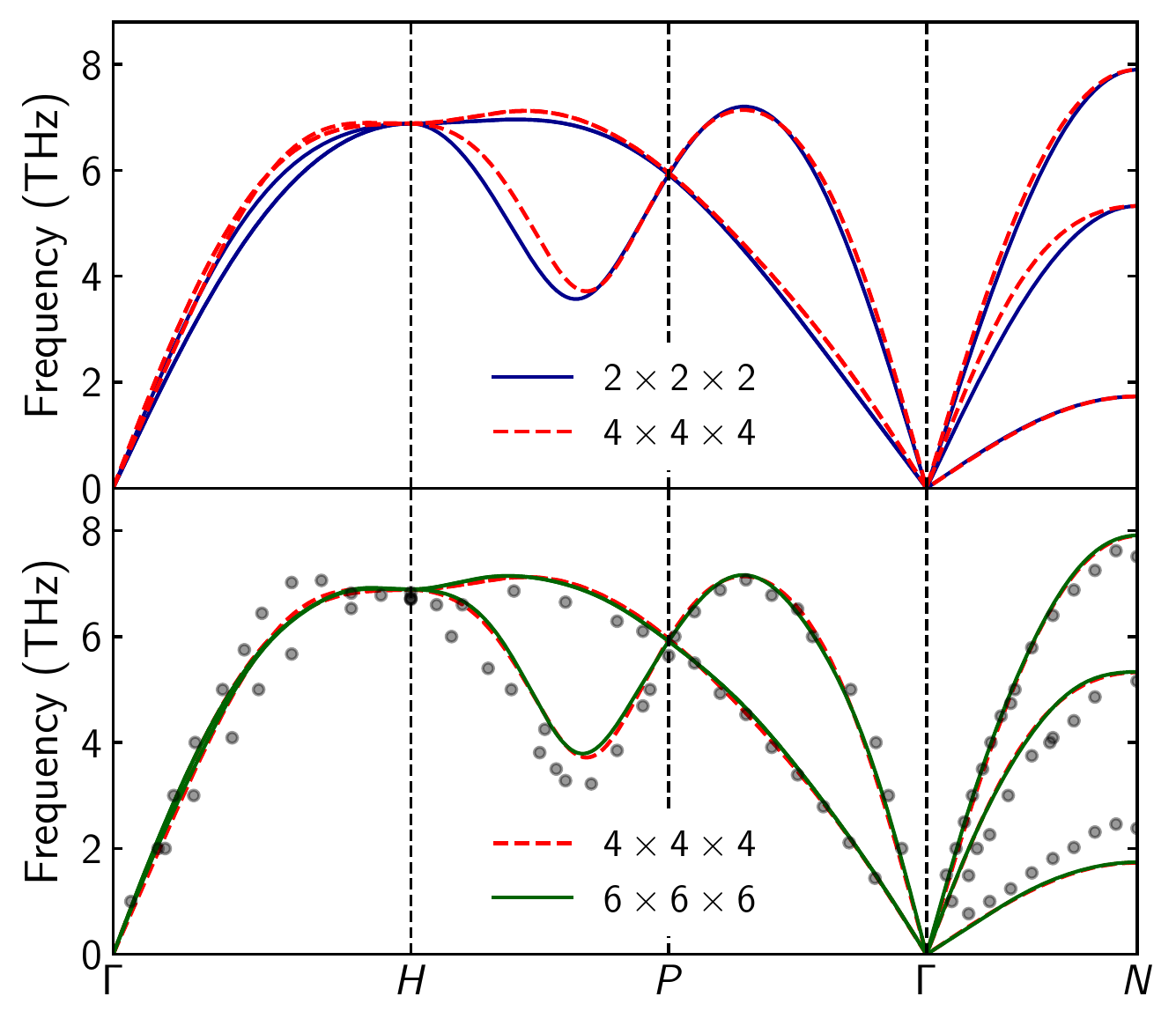}
    \caption{Convergence of the phonon dispersion of paramagnetic bcc $\delta$-Fe ($T=$~1740~K) with $\mathbf{q}$-point grid size. The grey dots correspond to the experimental data from Ref.~\cite{neuhaus2014}.}
    \label{fig:Fe_phon_PM}
\end{figure}

Here we compute the phonons for iron at temperatures of 1160~K and 1740~K at the experimental equilibrium volumes. The convergence of the phonon dispersion of paramagnetic $\delta$-Fe ($T=$~1740~K) is shown in Fig.~\ref{fig:Fe_phon_PM}. While previous work used a minimal $2\times2\times2$ $\mathbf{q}$-point grid~\cite{han2018}, there are small but visible differences between the results of a $2\times2\times2$ and a $4\times4\times4$ grid. The phonon dispersion is effectively converged with a $4\times4\times4$ grid, because the differences with the $6\times6\times6$ result are small. The advantage of nondiagonal supercells is clear with the largest grid; while for nondiagonal supercells, only supercells of up to six atoms are needed (scaling linearly with grid size $N$), the diagonal supercell would contain 216 atoms (scaling as $N^3$). For the case of iron the DFT+DMFT force calculations are remarkably robust; the difference between the force computed directly by DFT+DMFT ($\mathbf{F}_i$) and that required by the acoustic sum rule ($-\sum_{j\neq i} \mathbf{F}_j$) is smaller than 0.1\% (cf. Methods).

The agreement between the experimental data (grey dots in Fig.~\ref{fig:Fe_phon_PM}, Ref.~\cite{neuhaus2014}) and the calculations is very good, although there are small differences in the frequencies at certain $\mathbf{q}$-points, and the splitting of the branches along $\Gamma\to H$ is much smaller than in the experiment.

\begin{figure}[b]
    \centering
    \includegraphics[scale=0.625]{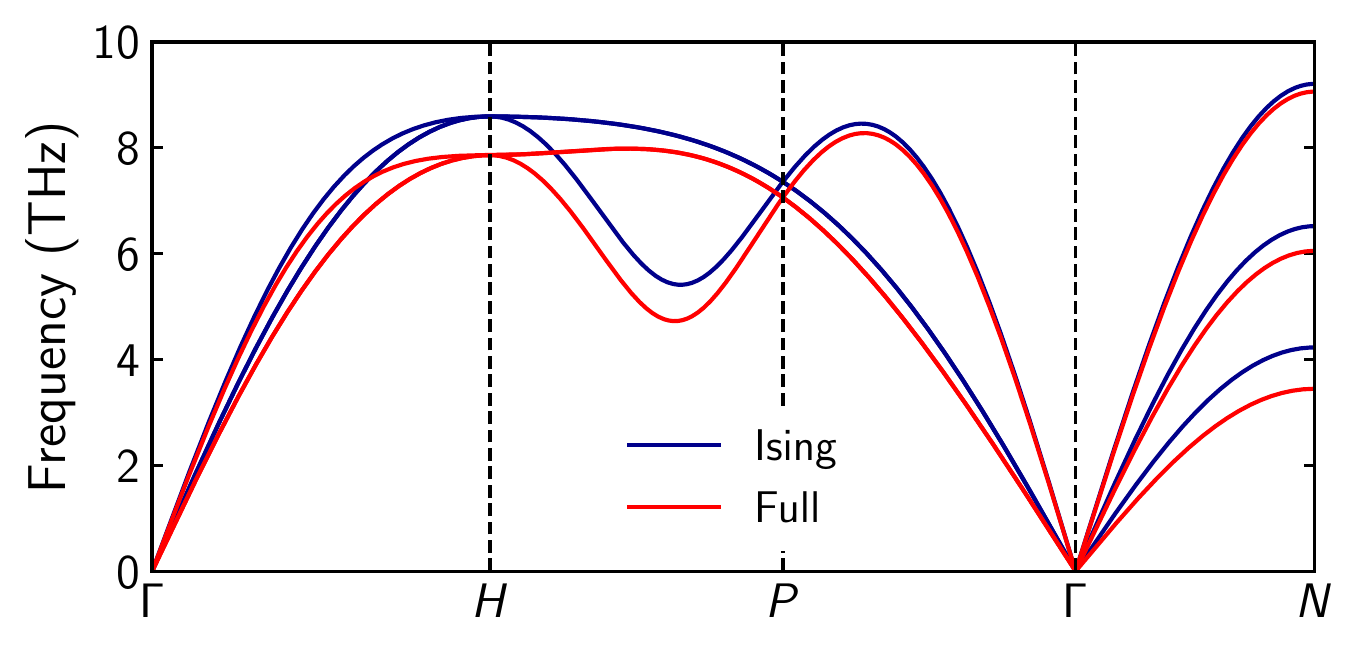}
    \caption{Phonon dispersion of ferromagnetic bcc $\alpha$-Fe ($T=$~1160~K) computed with a $2\times2\times2$ $\mathbf{q}$-point grid. The phonon dispersion shows clear differences between density-density (`Ising') and rotationally invariant (`Full') Coulomb interactions used in the impurity solver. With `Ising', the magnetic moment is larger than with `Full', and the phonons are consequently harder.}
    \label{fig:Fe_phon_FM}
\end{figure}

The phonon dispersion of the ferromagnetic $\alpha$-phase is shown in Fig.~\ref{fig:Fe_phon_FM} for a temperature of 1160~K. While the experimental Curie temperature $T_C$ is 1043~K, it is overestimated by DFT+DMFT calculations. This due to the fact that DMFT is a mean field theory and as such overestimates phase transition temperatures. Within DFT+DMFT, the transition temperature depends on the choice of Coulomb interaction in the impurity solver~\cite{anisimov2012,sakai2006}. The two options are the density-density only (`Ising') and rotationally invariant (`Full') Coulomb interaction. In the case of Fe, the choice of Coulomb interaction has an effect on the magnetic properties; while the Curie temperature with the Ising Coulomb interaction is  2500~K, using the Full interaction, the $T_C$ is 1550~K~\cite{han2018}. As a consequence, the magnetic moment for the same physical temperature of 1160~K is larger for Ising (2.38~$\mu_B$) than for Full (1.7~$\mu_B$). A comparison of the phonon dispersions at 1160~K calculated with Ising and Full Coulomb interactions demonstrates that the phonon frequencies with Ising are larger (Fig.~\ref{fig:Fe_phon_FM}). This difference is expected given the larger magnetic moment with Ising and the fact that the phonons in Fe soften with decreasing ferromagnetic order. For the paramagnetic case, no such difference between Ising and Full is observed and the phonon dispersions computed with the two methods are identical (not shown). This illustrates two important points: 1) it is consistent with the interpretation that the phonon softening is largely due to melting of the ferromagnetic order~\cite{han2018}, and 2) it suggests that phonons can be sensitive to the choice of Coulomb interaction. If this is the case, the approach used in Ref.~\cite{han2018} of scaling the physical temperature with respect to $T_C$ is appropriate.

\begin{figure}[!t]
    \centering
    \includegraphics[scale=0.625]{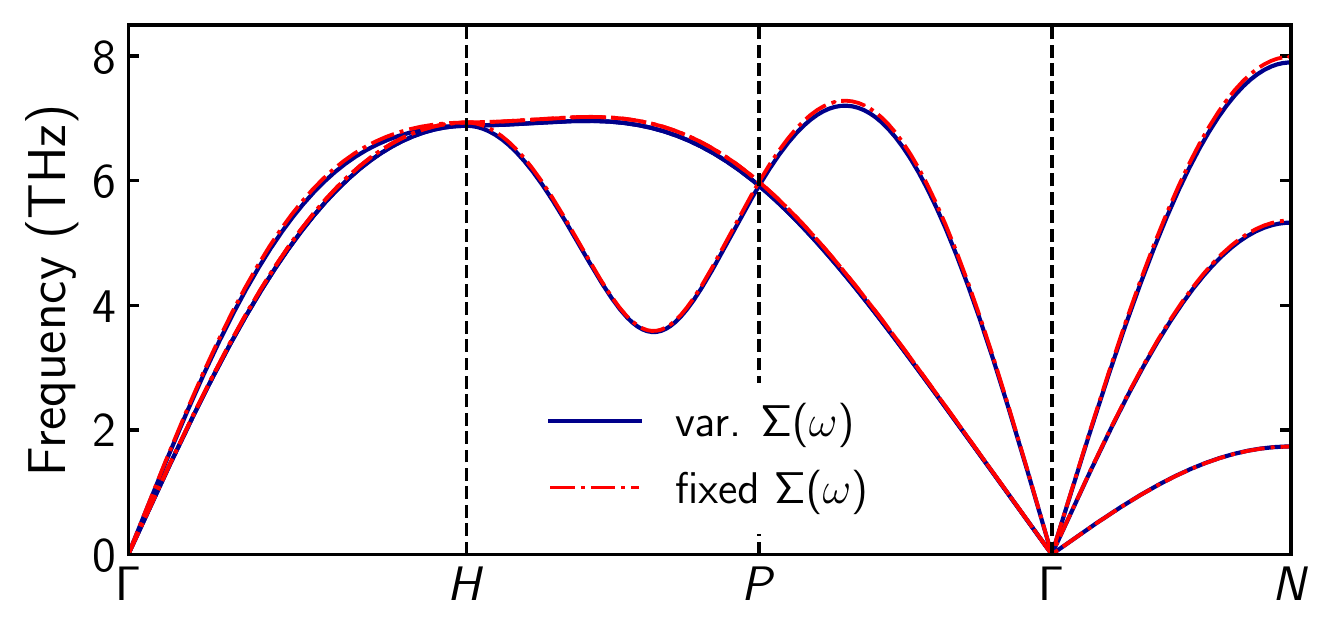}
    \caption{Phonon dispersion of paramagnetic bcc $\delta$-Fe ($T=$~1740~K, $2\times2\times2$ $\mathbf{q}$-point grid) with fixed and variable self-energy (see text). The results match to within 0.1 THz.}
    \label{fig:Fe_phon_PM_fixSE}
\end{figure}

While the use of non-diagonal supercells significantly speeds up the DFT+DMFT lattice dynamics calculations, it is still very expensive compared with DFT. The cost of the DFT+DMFT calculations increases with the number of atoms $N_{\mathrm{at}}$ in the unit cell as $aN_{\mathrm{at}}+bN_{\mathrm{at}}^3$, where the linear term is due to the solution of the quantum impurity problems, which dominates the cost of the DFT calculation (cubic term, small $b$) for reasonably sized systems~\cite{mandal2019}. The lattice dynamics calculations use very small atomic displacements (0.02~bohr) and it is worth checking how much the solution of the impurity problem is affected by the small changes in atomic positions. Fig.~\ref{fig:Fe_phon_PM_fixSE} shows the difference between phonon dispersions obtained by (1) solving the impurity problem separately for each correlated atom in the unit cell [variable $\Sigma(\omega)$], and 2) a calculation in which the self-energy for each correlated Fe atom is fixed to be equal to that of an Fe atom in the undisplaced equilibrium configuration [fixed $\Sigma(\omega)$]. The differences are very small; for example, at the $N$ point the phonon frequencies differ by at most 0.1 THz. Based on these results, it seems that fixing the self-energy is an excellent approximation for the case of iron. These differences are of the same magnitude as those that would be deemed acceptable when carrying out a convergence test of the phonon frequencies with respect to the basis set size ($RK_{\mathrm{max}}$), or the number of $\mathbf{k}$-points in the calculation. This suggests that the hybridisation and the impurity levels are not sensitive to the small displacements that are involved in the lattice dynamics calculation, and the impurity problem is hardly affected. Performing the calculations with a fixed self-energy offers massive benefits: a single DFT+DMFT calculations in the high-symmetry equilibrium configuration is sufficient to obtain $\Sigma(\omega)$. After that, the forces for structures with various displacements of atoms can be calculated at the cost of a DFT calculation, since all that is required is the calculation of the lattice Green's function with a fixed $\Sigma(\omega)$. Timing information shows that for the calculation of the $2\times2\times2$ $\mathbf{q}$-point grid shown in Fig.~\ref{fig:Fe_phon_PM_fixSE}, performing the calculations with a fixed self-energy is ten times faster. In general, the speedup depends on the ratio of the amount of time spent in the impurity solver to the time spent in other steps of the calculation.

It is important to examine the validity of fixing the self-energy in each individual case. As shown in the following sections, for the materials studied in this work, a fixed self-energy is an excellent approximation. Put differently, this means that the change in the self-energy with respect to position is small and therefore the two-particle vertex function is not important for computing forces. However, one might expect that this approximation breaks down when the material is close to a phase transition. At present, we recommend testing this approximation in particular cases that are under consideration. In the following, we have tested it for the materials studied by comparing at least the $\Gamma$-point phonon frequencies with a fixed self-energy to those of a non-approximated calculation.

\subsection{NiO and MnO}
NiO and MnO are antiferromagnetic insulators with Neel temperatures $T_N$ of 525~K and 116~K, respectively. Above $T_N$, the compounds are paramagnetic insulators. Magnetic ordering in NiO and MnO induces a change in crystal symmetry; while the high temperature phases are cubic, the low temperature AFM phases are rhombohedral. The phonon spectra of both NiO and MnO do not depend sensitively on the presence of long-range magnetic order~\cite{reichardt1975}, but the change in crystal symmetry leads to small changes in the phonon frequencies due to magnetic anisotropy~\cite{massidda1999,wang2010,chung2003}. One of the advantages of using DFT+DMFT over DFT for lattice dynamics calculations of NiO and MnO is the ability to simulate the paramagnetic regime directly. We therefore chose to perform the calculations for the paramagnetic regime at room temperature. MnO is in fact paramagnetic at room temperature, and for NiO this is a commonly used simplification.

\begin{figure}[b]
    \centering
    \includegraphics[scale=0.63]{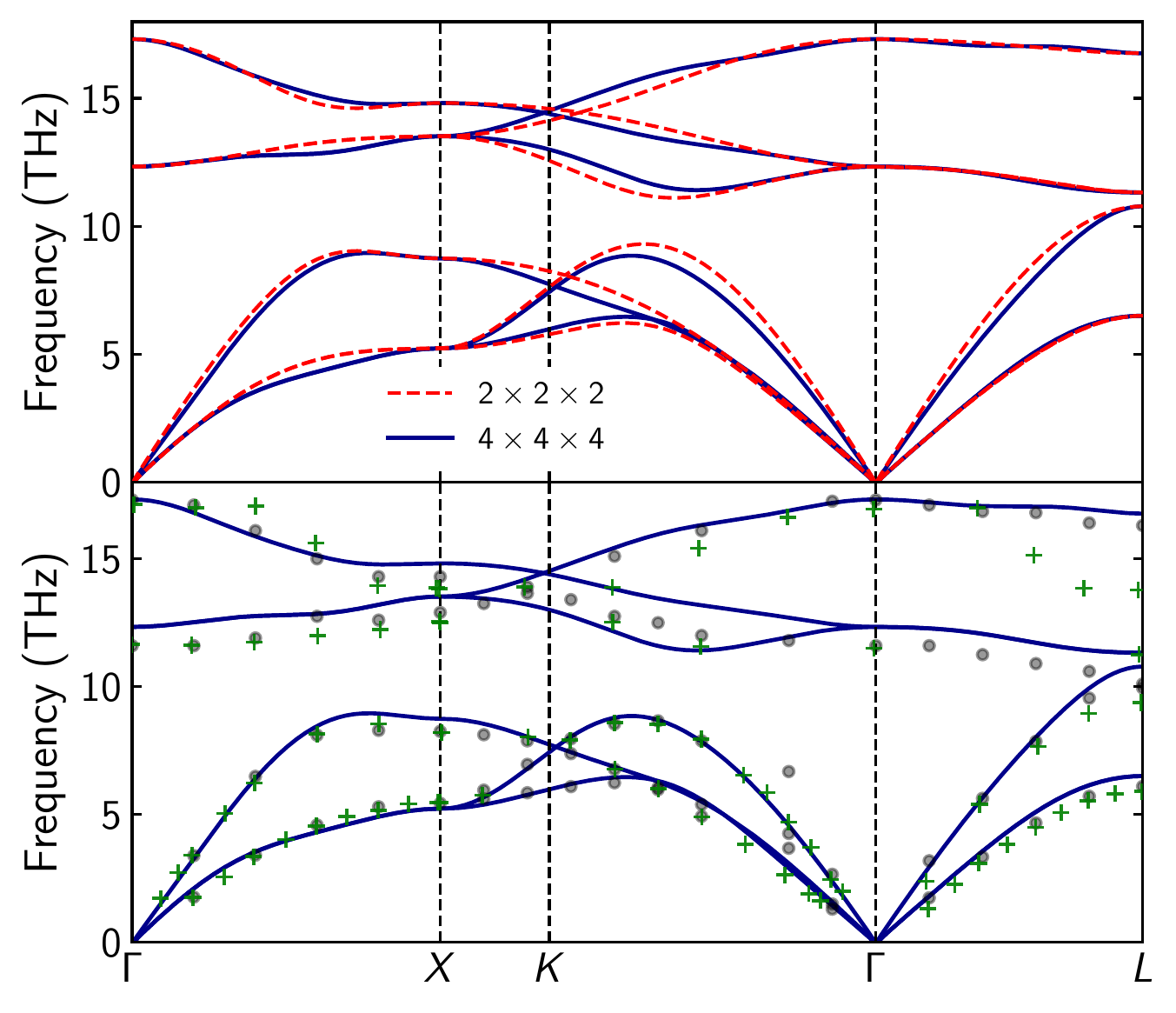}
    \caption{Convergence of the phonon dispersion of NiO with the size of the $\mathbf{q}$-point grid. The grey dots correspond to the data of Reichardt~\textit{et~al.}~\cite{reichardt1975}, and the green crosses to the data of Coy~\textit{et~al.}~\cite{coy1976}.}
    \label{fig:NiO_phon}
\end{figure}

The cubic phase is also convenient because it simplifies including LO-TO splitting for MnO and NiO. Since it is currently not possible to calculate Born effective charge tensors with DFT+DMFT, the LO-TO splitting has to be calculated using elongated supercells that represent $\mathbf{q}$-points close to $\Gamma$. It is much easier to do this for the cubic paramagnetic phases than for the AFM phases. Unfortunately, elongated supercells [e.g. corresponding to $\mathbf{q}=(0,0,\frac{1}{8})$] led to problems with the DFT+DMFT force calculations. Specifically for the binary crystal structures NiO and MnO, the nondiagonal supercells showed non-zero forces on atoms even in the high-symmetry equilibrium configuration. The problems became more severe with larger cell sizes, increasingly unequal lattice parameters, and large deviations of unit cell angles from 90$^\circ$. While the computational expense of a $6\times6\times6$ or $8\times8\times8$ $\mathbf{q}$-point grid would have been manageable, the systematic issues with the forces prevented the use of larger grids. We note that these issues were not due to statistical noise in the impurity solver and seem to leave room for improvement of the implementation. These problems were not encountered in the case of Fe, and therefore seem to be related to having two atomic species present in the cell, one of which is being treated as correlated while the other is not. Note that while this has an effect on lattice dynamics calculations, which involve small atomic displacements, the force implementation works very well for structural optimisation of correlated materials~\cite{haule2017,stanislavchuk2020,pascut2020}. Since we were unable to extract the LO and TO mode frequencies, we instead use values of $Z^*$ and $\epsilon_\infty$ from Ref.~\cite{savrasov2003} or experiment~\cite{gielisse1965,plendl1969} for the LO-TO splitting. While unsatisfactory, there is currently no other method of including the LO-TO splitting using finite differences.

\begin{figure}[!htb]
    \centering
    \includegraphics[scale=0.63]{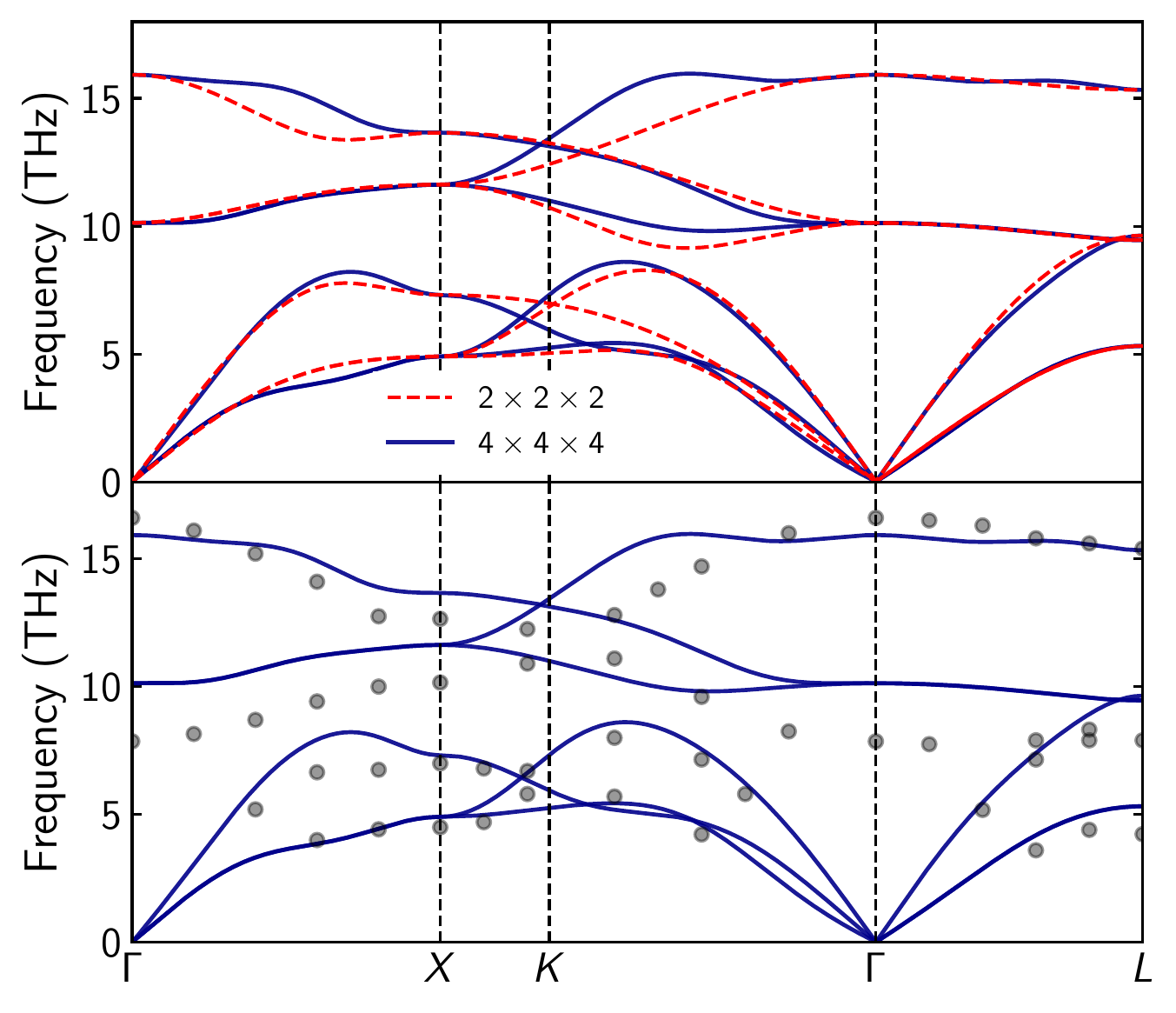}
    \caption{Convergence of the phonon dispersion of MnO with the size of the $\mathbf{q}$-point grid. The grey dots correspond to the data of Wagner \textit{et al.}~\cite{wagner1976}.}
    \label{fig:MnO_phon}
\end{figure}

Phonon dispersions for NiO with different $\mathbf{q}$-point grid sizes are compared in Fig.~\ref{fig:NiO_phon}. The differences between the $2\times2\times2$ and $4\times4\times4$ $\mathbf{q}$-point grids are larger than for the case of Fe, which can be attributed to the stronger screening in the metal. Larger grids are not accessible due to stability issues with the force implementation. For NiO, non-diagonal supercells allow us to access a grid of $4\times4\times4$ with supercells that contain at most 4 primitive cells (8 atoms). A diagonal supercell would contain 64 primitive cells (128 atoms), which would be much more computationally expensive. The DFT+DMFT phonon dispersions are compared to two sets of experimental data from Refs.~\cite{reichardt1975,coy1976} in Fig.~\ref{fig:NiO_phon}. The agreement is generally good, and is on the order of the differences between the two experiments. The acoustic branches show an overall better agreement with the experimental data than the optical branches. Using the self-energy of the equilibrium configuration for the lattice dynamics calculation is also an excellent approximation for NiO; for example, the TO mode frequency at $\Gamma$ changes by less than 0.01 THz. Depending on the shape of the nondiagonal supercell the differences between the results obtained from a fixed vs variable self-energy can be larger, but this is likely due to the issues with the DFT+DMFT force calculations discussed above.

Phonon dispersions for MnO with grids of sizes $2\times2\times2$ and $4\times4\times4$ are shown in Fig.~\ref{fig:MnO_phon}. As for NiO, there are significant differences in the phonon dispersions obtained with a larger $\mathbf{q}$-point grid. The DFT+DMFT results for the $4\times4\times4$ grid are compared to the experimental data of Wagner~\textit{et~al.}~\cite{wagner1976}. For MnO, the agreement with experiment is not as good as for NiO. This is mostly due to a difference of 2.3~THz between the experimental and calculated TO phonon mode frequency at $\Gamma$.

The vibrational properties of MnO and NiO have been studied previously by DFT with different functionals, including DFT+U and hybrid functionals~\cite{wdowik2009,floris2011,linnera2019}. The study by Linnera \textit{et al.}~\cite{linnera2019} used hybrid functionals and obtained good agreement with the experimental phonon spectrum of NiO, but a strong underestimation of the optical phonons at $\Gamma$ for MnO. DFT+U was used in Refs.~\cite{wdowik2009,floris2011}, obtaining good agreement with experimental phonon frequencies for MnO when choosing an appropriate $U$ value. We have tested $U$ values in the range 8--10~eV in DFT+DMFT calculations for MnO, but this did not improve agreement with the experimental phonon frequencies (although tuning to much lower $U$ values might). Given the charge-transfer insulating nature of MnO an insensitivity to the precise $U$ value is expected.

\subsection{\ce{SrVO3}}
The perovskite \ce{SrVO3} is often cited as a textbook example of a strongly correlated metal. The vanadium atom nominally has a $d^1$ configuration with a single electron in its $t_{2g}$ subshell. The \ce{SrVO3} spectral function shows a well-established three-peak structure, with a quasiparticle peak around the Fermi level, and pronounced lower and upper Hubbard bands below and above~\cite{sekiyama2004,haule2014}. We have calculated the phonons of \ce{SrVO3} at $T=293$~K with DFT+DMFT to assess the effect of strong correlations on the vibrational properties of the material. Note that due to the metallic nature of the material there is no LO-TO splitting.

\begin{figure}[!htb]
    \centering
    \includegraphics[scale=0.625]{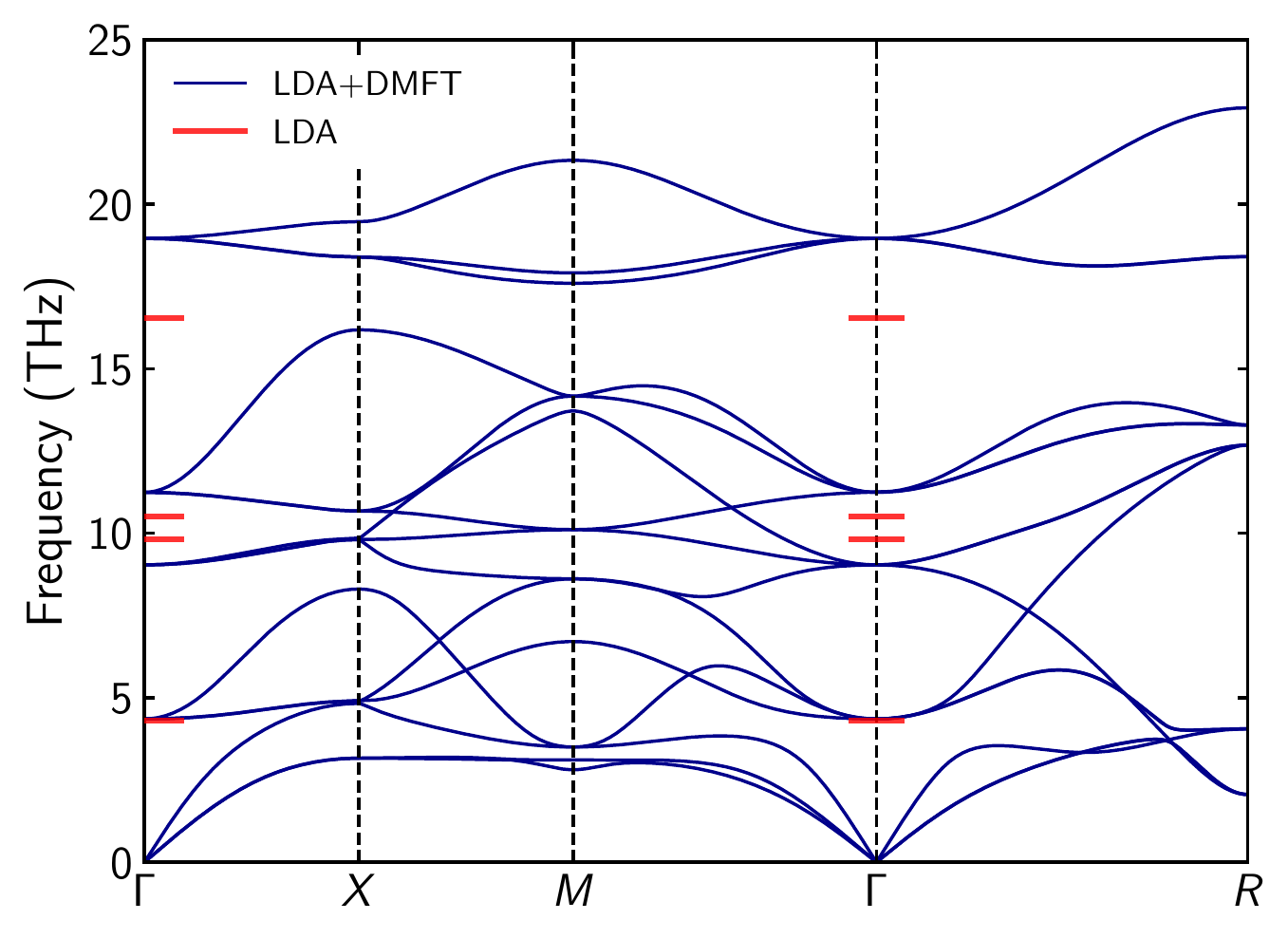}
    \caption{Phonon dispersion of \ce{SrVO3} ($T=293$~K) computed with a $2\times2\times2$ $\mathbf{q}$-point grid, at the LDA+DMFT level of theory. Frequencies of $\Gamma$ point phonons calculated with LDA are marked in red.}
    \label{fig:SrVO3_phon}
\end{figure}

For \ce{SrVO3}, the violation of the acoustic sum rule is much more severe than for Fe, NiO, or MnO. The condition of Eqn.~\ref{eqn:ASR} is satisfied only to within 5--7\% in the worst cases, which is a significantly larger violation than for the previous materials. For displacements that involve the correlated vanadium atom and its nearest neighbours the violation of the acoustic sum rule is worse than for displacements that only involve Sr. This issue likely arises due to contributions to the forces from terms that depend on the correlated subspace. In the case of Fe, all atoms are equivalent and this term cancels, while for NiO, MnO, and \ce{SrVO3} these terms cannot cancel because of the presence of different atomic species. Enforcing the sum rule is essential to obtain useful results.

As for the other materials, reusing the self-energy of the equilibrium configuration for the lattice dynamics calculation is an excellent approximation. For the modes at $\Gamma$, the frequencies computed with a fixed vs variable self-energy differ by less than 0.01 THz. We therefore performed the calculations of a $2\times2\times2$ $\mathbf{q}$-point grid for \ce{SrVO3} with a fixed self-energy. The resulting LDA+DMFT phonon dispersion is shown in Fig.~\ref{fig:SrVO3_phon} and compared to the LDA $\Gamma$-point phonon frequencies, since experimental data for the vibrational properties of \ce{SrVO3} is not available. The results confirm the dynamical stability of the \ce{SrVO3} perovskite structure at the LDA+DMFT level of theory. Focusing on the $\Gamma$ point, there are five threefold degenerate phonon modes. The frequencies of the modes are renormalised by correlation effects by different amounts. The frequency of the highest mode changes the most, while the frequency of the lowest mode, dominated by Sr moving against an almost rigid \ce{VO6} octahedron, is the same for LDA and LDA+DMFT. This is expected since the correlated atom and its nearest neighbours do not change their relative positions, and confirms the internal consistency of the LDA+DMFT phonon calculations; if a phonon mode does not involve the motion of the correlated atoms or their direct neighbors, the frequency should be unchanged from the DFT value. On the other hand, if a mode features large changes in the relative positions of a correlated atom and its nearest neighbors, correlation effects can be expected to strongly impact the frequency of that mode. This is the case for the highest frequency $\Gamma$-point phonon, which is more strongly affected by correlation effects because the vanadium and oxygen atoms move relative to each other. These observations indicate the most useful applications of DMFT phonon calculations: phonon effects that depend on correlations and temperature due to the involvement of correlated atoms in the atomic motion.

\section{Conclusion}\label{sec:Conclusion}
In this paper, we have described a method to efficiently compute phonons in correlated materials using a DFT+DMFT approach. The method combines a robust DFT+DMFT force implementation with the use of nondiagonal supercells for finite difference lattice dynamics calculations. We have calculated phonons of multiple different correlated materials, including metals and insulators, elemental, binary and ternary crystals. The efficiency of the method allowed us to access $\mathbf{q}$-point grids of very large size. The agreement between the calculations and available experimental data is generally good. Based on our tests, the self-energy obtained from a DFT+DMFT calculation of the equilibrium configuration is accurate enough for lattice dynamics calculations, which eliminates the need to solve a large number of impurities for configurations with displaced atoms. Finally, we have discussed some issues with the DFT+DMFT force implementation that should be solved to make the calculation of phonons using finite differences more robust.

There are many problems in condensed matter physics of strongly correlated materials that would benefit from an elucidation of the phonons with DFT+DMFT. Cases that come to mind are the phase diagram of $f$-elements such as cerium and uranium, and the metal-insulator transitions in vanadate materials. The method should be especially useful to evaluate phonons close to phase transitions to clarify the role they play in correlated materials, and for the interpretation of temperature dependent diffuse scattering across structural phase transitions~\cite{gutmann2015,girard2019}. More generally, finite difference approaches for lattice dynamics will also be useful for calculations of electron-phonon coupling and a variety of other phenomena in strongly correlated materials that depend on atomic vibrations.

\begin{acknowledgments}
C.P.K. thanks the Winton Programme for the Physics of Sustainability and EPSRC for financial support. K. H. acknowledges funding by the National Science Foundation through grant NSF-DMR 1709229. G.L.P. was supported by contract no.\ 18PFE/16.10.2018 funded by the Ministry of Research and Innovation of Romania within Program 1 - Development of National Research and Development System, Subprogram 1.2 - Institutional Performance-RDI Excellence Funding Projects. B.M. acknowledges support from the Gianna Angelopoulos Programme for Science, Technology, and Innovation, and from the Winton Programme for the Physics of Sustainability. We acknowledge the use of the Cambridge Service for Data Driven Discovery (CSD3) operated by the University of Cambridge Research Computing Service, funded by the EPSRC (EP/P020259/1), and DiRAC funding from the STFC.
\end{acknowledgments}

\bibliography{refs}

\end{document}